\documentclass[sigconf]{acmart}

\AtBeginDocument{%
  \providecommand\BibTeX{{%
    \normalfont B\kern-0.5em{\scshape i\kern-0.25em b}\kern-0.8em\TeX}}}

\makeatletter
\@namedef{ver@lineno.sty}{9999/12/31}
\@namedef{opt@lineno.sty}{}
\makeatother

\usepackage{graphicx}
\usepackage{balance}
\usepackage{xcolor}
\usepackage{xargs}                      
\usepackage{xparse}
\usepackage[colorinlistoftodos,prependcaption]{todonotes}
\definecolor{atomictangerine}{rgb}{1.0, 0.6, 0.4}
\definecolor{columbiablue}{rgb}{0.61, 0.87, 1.0}
\definecolor{lava}{rgb}{0.81, 0.06, 0.13}
\usepackage[frozencache,cachedir=listings]{minted}
\usepackage[most, minted]{tcolorbox}

\usepackage{subcaption}

\usepackage{stfloats}

\newcommand{\dquote}[1]{''#1''}
\newcommand{\squote}[1]{'#1'}

\newcommandx{\unsure}[2][1=]{\todo[linecolor=atomictangerine,backgroundcolor=atomictangerine!25,bordercolor=atomictangerine,#1]{#2}}
\newcommandx{\fix}[2][1=]{\todo[linecolor=lava,backgroundcolor=lava!25,bordercolor=lava,#1]{#2}}
\newcommandx{\info}[2][1=]{\todo[linecolor=columbiablue,backgroundcolor=columbiablue!25,bordercolor=columbiablue,#1]{#2}}
\newcommandx{\improvement}[2][1=]{\todo[linecolor=Plum,backgroundcolor=Plum!25,bordercolor=Plum,#1]{#2}}
\newtcblisting{openhps}{%
	listing engine=minted,
	colback=gray!5,			
	minted language=minted/lexers/typescript.py:TypeScriptLexer -x,
	minted style=vs,
	listing only,
	enhanced,
	boxrule=1pt,			
	colframe=gray!20,		
	minted options = {
		linenos, 			
		breaklines=false,	
		fontsize=\footnotesize, 
		numbersep=2mm,		
		tabsize=2,			
		stripall=true		
	},
	overlay={
		\begin{tcbclipinterior}
			\fill[gray!20] (frame.south west) rectangle ([xshift=4mm]frame.north west);
		\end{tcbclipinterior}
	},
	before upper={
		\footnotesize
		\mintinline{typescript}{/* @openhps/core | version 0.2.0 */}
	}
}

\setcopyright{none}

\makeatletter
\renewcommand\@formatdoi[1]{\ignorespaces}

\begin{document}

\copyrightyear{This article has been peer reviewed and presented at the 5th International Workshop on (Document) Changes: Modeling, Detection, Storage and Visualization (DChanges~2018) in Halifax, Canada}
\acmConference[DChanges 2018]{ACM Symposium on Document
Engineering 2018}{August 28, 2018}{Halifax, Canada}
\acmISBN{}
\acmPrice{}

\title{A Metamodel and Prototype for Fluid Document Formats}


\author{Ahmed A.O.~Tayeh}
\orcid{https://orcid.org/0000-0002-3576-6802}
\affiliation{%
	\institution{WISE Lab\\
		Vrije Universiteit Brussel}
	\streetaddress{Pleinlaan 2}
	\city{1050 Brussels}
	\country{Belgium}
}
\email{atayeh@vub.be}

\author{Bruno Dumas}
\orcid{ https://orcid.org/
0000-0001-5302-4303}
\affiliation{%
	\institution{PReCISE, NADI\\
		University of Namur}
	\streetaddress{Rue de Bruxelles 61}
	\city{5000 Namur}
	\country{Belgium}
}
\email{bruno.dumas@unamur.be}

\author{Beat Signer}
\orcid{https://orcid.org/0000-0001-9916-0837}
\affiliation{%
	\institution{WISE Lab\\
		Vrije Universiteit Brussel}
	\streetaddress{Pleinlaan 2}
	\city{1050 Brussels}
	\country{Belgium}
}
\email{bsigner@vub.be}

\begin{abstract}
  With the transformation of computing from personal computers to the Internet, document formats have also seen some changes over the years. Future document formats are likely going to adapt to the emerging needs of ubiquitous computing, where information processing is embedded in everyday activities and objects. While most existing document formats have originally been a digital emulation of paper documents, over the years they have been enriched with additional digital features. These features were mainly incorporated to take advantage of the new functionality offered by the devices on which the documents are accessed. With the advent of ubiquitous computing, document formats seem to be facing the next evolutionary step. They will have to adapt to novel mobile devices, innovative interaction modalities, the distribution over multiple devices as well as heterogeneous input sources. This adaptation to the age of ubiquitous computing asks for several new document features. We outline a roadmap towards future fluid document representations for ubiquitous information environments. Based on the resource-selector-link~(RSL) hypermedia metamodel---a general hypermedia metamodel supporting distribution, user rights management and content adaptation---we developed a metamodel for fluid document formats and the corresponding online text editor for fluid documents.

\end{abstract}


\maketitle

\section{Introduction}
We are currently at the brink of the age of ubiquitous computing, where any object is going to be a computer, each sensor a server and every human being at the very core of a cloud of devices. Document formats have accompanied the transformation from \dquote{personal} computing towards ubiquitous computing. However, often the recent evolution of computing has only been endured rather than followed by most document formats. A good example is how document formats nowadays must adapt to the small screen size of smartphones. Even though current smartphones are equipped with high-resolution screens, the actual size of the device forced content providers to find ways to adapt to the content of documents that were originally intended to be printed on A4 or letter-size paper. In the near future, it is likely that the range of devices on which documents are supposed to be displayed will become even more diversified. Apart from the screen size, other features including significant variations in memory capacity or the available bandwidth will have to be considered as well. Further, documents might be distributed across multiple devices or servers and flow between them with or without human intervention.

To address the multitude of challenges imposed by the emerging age of ubiquitous computing, document formats will have to take into consideration a range of digital features. We believe that advanced linking is needed since content is going to be distributed and split across different devices. In fact, the unidirectional linking we know from the Web has to evolve into bidirectional or even multidirectional linking. With the potential scattering of resources with multiple versions and identifiers, some more adequate support for transclusion~\cite{Krottmaier2001} enabling the reuse of parts of documents might eventually be needed. Another important feature that should be considered in future document formats is versioning with a full list of modifications that have been applied to a document, as well as information about the users who made these modifications. Such a versioning feature combined with multidirectional links and transclusion will provide access to the complete history of changes and versions of a given document. Another important feature that should be taken into account is digital rights management. The inclusion of user rights management at the document level or even at the level of parts of a document will enable us to address authorisation in a deeply interconnected digital world. Finally, beyond the adaptation of a document's representation resulting in different output, the adaptation of input modalities will also gain importance. The knowledge on how to interact with content might become as important as knowing how to present it.

In the next section, we discuss the current support of the aforementioned digital features in existing document formats. We then outline a roadmap towards future fluid document formats and discuss a metamodel and prototype for future fluid document formats. Finally, we outline future work and provide some conclusions.

\section{Document Support for Advanced Digital Features}
Existing document formats provide good support for \emph{embedded unidirectional hyperlinks} as well as for some limited forms of transclusion, such as the inclusion of images in HTML documents~\cite{Krottmaier2001}. However, support for more advanced links and full transclusion is limited. In more detail, bi- and multi-directional links have been somewhat explored in the XML Linking Language~(XLink)~\cite{Christensen2002}. Full transclusion has been explored by XML technologies, namely the conjunction between XLink and XPointer\footnote{http://www.w3.org/TR/xptr-xpointer/}. Even though more recent document formats such as the Office~Open~XML~(OOXML)\footnote{ttps://www.openoffice.org/xml/} and OpenDocument\footnote{ttp://opendocumentformat.org} standards are XML-based document formats, they only support the unidirectional \emph{simple links} (similar to HTML~links) of the XLink specification but do not offer the more advanced \emph{extended links}. While OOXML supports a restricted version of transclusion (e.g.~for the inclusion of images), the OpenDocument format exploits the XLink \squote{show} behaviour attribute to support full transclusion.

Versioning also has limited support in most existing document formats. It is supported in the OpenDocument and OOXML formats via the track changes mechanism introduced by Microsoft's Office suite, which forced both formats to include fine-grained modification tracking. The Standard Generalized Markup Language~(SGML) contains a set of elements and attributes enabling a top-down or version~ID-centred form of versioning, in contrast to the bottom-up or atomic change-centred approach offered by OpenDocument and OOXML. User rights management has been approached only from the digital rights management perspective by EPUB~3\footnote{http://idpf.org/epub/30} and the Portable Document Format~(PDF). Even if this represents a form of granting authorisation, no functionality for user profile management is provided. Finally, for content adaptation we can differentiate between the adaptation of output and the adaptation of input. With respect to output adaptation, HTML as well as XHTML\footnote{http://www.w3.org/TR/xhtml11/} documents have the potential to adapt their layout to different devices, in particular when used in combination with Cascading Style Sheets~(CSS) and also EPUB~3 provides numerous forms of content adaptation.

\section{Towards Future Document Formats}
Towards the \dquote{ideal} future document representation, we believe that multiple tracks can be followed. The first track is to create a brand-new document format considering a full set of requirements dedicated to ubiquitous computing. Even if this track first appears as the most evident one, it is also the most prone to failure. Indeed, introducing a new document format without clear motivation and strong support would almost certainly end up in poor acceptance. Most \dquote{new} document formats are in fact extensions of one or multiple standards to guarantee easier acceptance. This means that they also inherit the potential weaknesses of these ancestor formats. The track followed by the new formats such as EPUB, OpenDocument or OOXML has been to package multiple sub-documents in already well-established formats. A second track that could be followed is to extend existing document formats with the features needed for ubiquitous computing. In fact, this track will almost certainly be followed by several formats. Obviously, this will primarily worsen the complexity of these formats, while answering the needed features one step at a time.

We believe that there exists a third track that does not consist of adding a new layer of complexity on top of existing document formats, but rather below them in an intermediary layer. With the adaptation of document formats to the emerging age of ubiquitous computing, documents will more often be stored on servers and in databases and no longer only in local file systems. This will enable remote access to these documents through RESTful interfaces or Web Services. The Ubiquitous Computing and Internet of Things research communities are examining middleware architectures dealing with the storage, localisation and addressing of data in heavily distributed and heterogeneous environments~\cite{Guinard2011,Presser2009}. Documents will in turn form part of the Internet of Things either as complete entities, or different snippets of a document might be distributed over multiple locations. Note that document formats such as EPUB~3, OpenDocument or OOXML already consist of a package that can be decomposed into multiple sub-documents.

Based on some of the concepts presented earlier, Boyer~et~al.~\cite{Boyer2009} already demonstrated the interest in decomposing documents in OpenDocument format into interactive web documents which can then be addressed through a RESTful interface. Krottmaier~et~al.~\cite{Krottmaier2001} proposed a similar server-based approach for implementing transclusions. We also think that such a middleware-based approach should be further explored and exploited. Indeed, by embedding metadata for individual pieces of information, advanced features such as the ones discussed in the previous section could be supported. Bidirectional and multidirectional links should be first-class citizens and can be used to combine different subparts of a document as well as to enable the transclusion of parts of documents. This mechanism has already been adopted to realise an associative file system~ \cite{Signer2010}. As each piece of information would be identified by a unique \dquote{fingerprint} identifier generated based on its content, transclusion as originally proposed by Ted Nelson could be supported. M\"{u}ller~et~al.~\cite{Muller2010} already demonstrated how integrating versioning as a layer between the file system and documents offers multiple advantages. Note that user profiles and rights could also be embedded at this level. Finally, as documents would consist of a collection of information snippets rather than one monolithic file, the adaptation of content could also be supported.

Such a middleware for fluid document formats could be realised based on a metamodel offering features for the distribution, versioning, user rights management and adaptation of content. Different document formats might effectively be modelled on top of the metamodel and at different levels of granularity. For example, a document format tightly linked to the metamodel would map its elements to individual components of the metamodel. With such an approach, all features supported by the metamodel would also be supported by the document format. For document formats not directly supporting the metamodel, a document would be embedded in a single component of the metamodel, still allowing it to support some of the features of the underlying metamodel. Finally, any information managed by the metamodel could be stored either in the middleware or as separate metadata in addition to the document content.

\begin{figure}[htb]
\centering
\includegraphics[width=0.83\columnwidth]{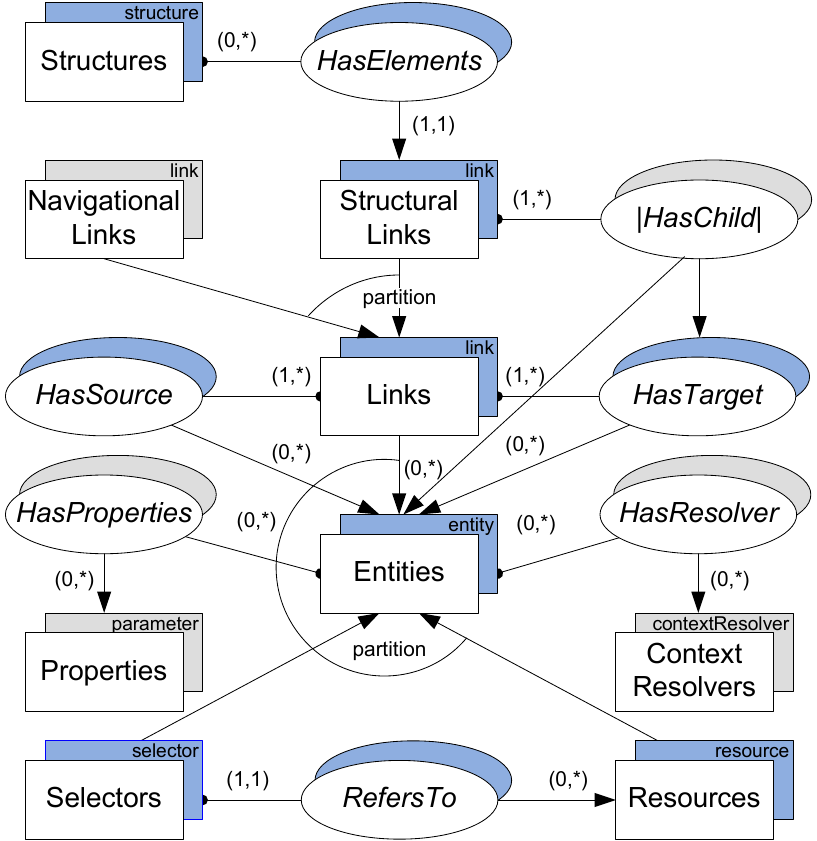}
\caption{Core RSL components}
\label{rsl}
\end{figure}

The metamodel of such a middleware for fluid document formats might be based on the resource-selector-link~(RSL) hypermedia metamodel by Signer and Norrie~\cite{Signer2007}, whose core components are illustrated in Figure~\ref{rsl}. The RSL hypermedia metamodel is based on the principle of linking arbitrary entities, whereby an entity can either be a \emph{resource}, a \emph{selector} or a \emph{link}. A resource defines a media type such as a text, a video or a complete document. A selector always must be attached to a resource and is used to address parts of a resource. Finally, a link can be a one-to-one, one-to-many, many-to-one or many-to-many bidirectional association between arbitrary entities. In addition, the RSL~hypermedia metamodel offers features such as user rights management, content distribution, content adaptation and the support for \emph{structural links} which can be used to compose a new entity (e.g.~a document) out of existing entities. Over the last decade, the RSL~hypermedia metamodel has proven its flexibility, generality and extensibility, and served as a basis for the implementation of a variety of hypermedia and cross-media solutions~\cite{Signer2005,Tayeh2014,Tayeh2018}.

\section{Fluid Document Format Prototype}
We have exploited the RSL~hypermedia metamodel to model future fluid document formats. Fortunately, no extensions of RSL were required since the RSL hypermedia metamodel already provides all the necessary components for composing and modelling fluid documents. As mentioned earlier, RSL offers the concept of structural links which can be used for composing new entities out of existing ones. A newly composed entity---which might be a document in our case---is built from the specific resources supported by the metamodel (e.g.~text or multimedia content) and from already existing composite structures. In the context of document formats, a newly composed structure might, for instance, be a section, a chapter or a book. A document that is completely mapped to RSL~resources, selectors and links benefits from all the features already provided by the RSL~hypermedia metamodel, including the support of bi- and multidirectional hyperlinks, user rights management, content adaptation based on \emph{context resolvers} as well as cross-media transclusion. The RSL~hypermedia metamodel further already supports the distribution of content through bi- and multidirectional structural hyperlinks.

\begin{figure}[htb]
\centering
\includegraphics[width=\columnwidth]{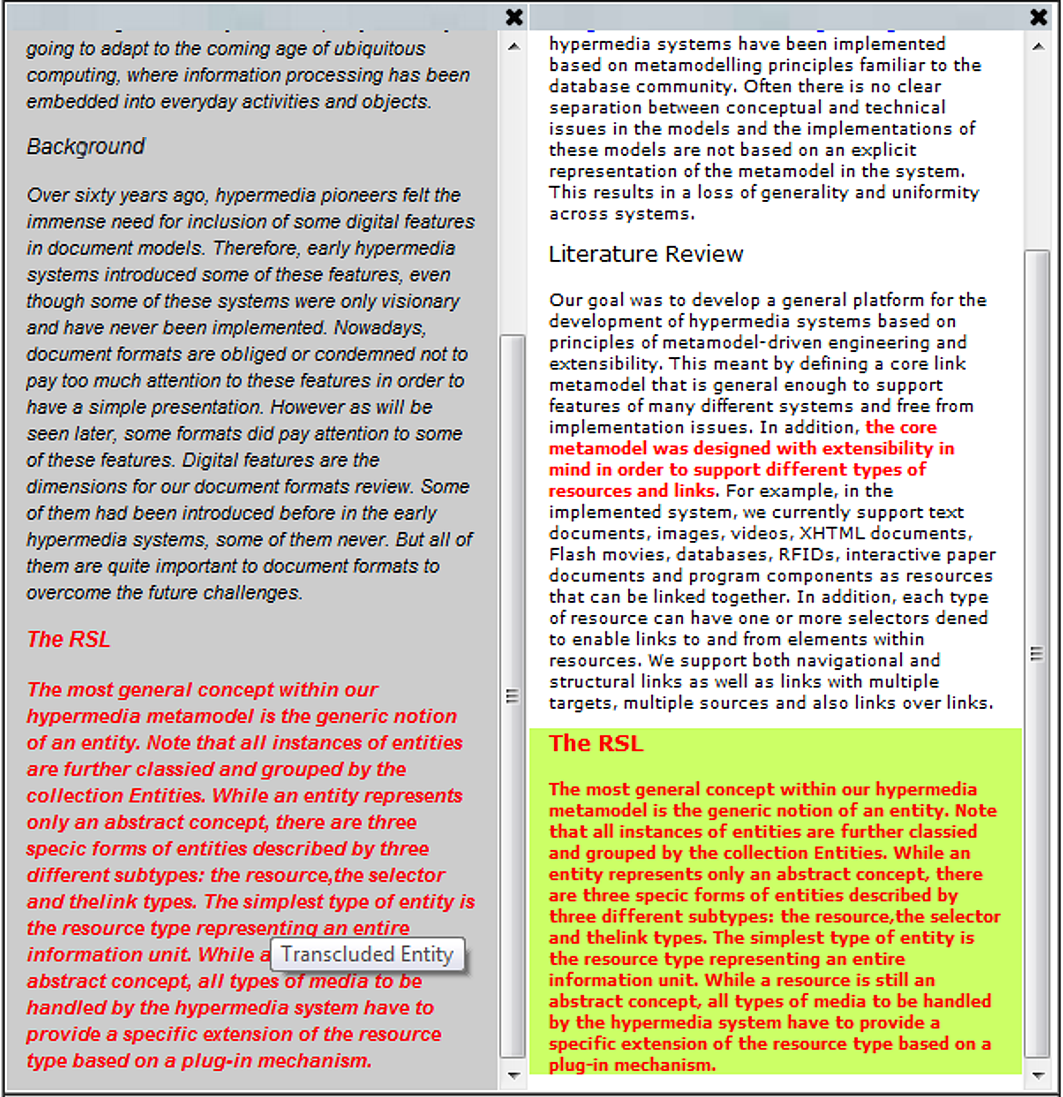}
\caption{Content transclusion in fluid documents}
\label{transc}
\end{figure}

In order to prove that fluid document formats modelled and implemented based on the RSL~hypermedia metamodel profit from the features offered by the metamodel, we have developed a proof-of-concept prototype for fluid document formats. The prototype enables the distribution of parts of documents that can be accessed and composed via a RESTful~API. Users can visualise documents side by side which facilitates the creation of bi- and multidirectional hyperlinks as well as the transclusion of content as highlighted in the screenshot of our fluid document formats prototype in Figure~\ref{transc}. Based on the user access rights defined for different document parts, the prototype highlights the document parts a user has access to. A document therefore might be visualised differently for each user having access to the document. Last but not least, we have exploited the RSL \emph{context resolver} and \emph{property} components to model user preferences for specific parts of a document, enabling the adaptation of output based on individual user preferences.

\section{Conclusion}
Our research proposes potential future directions for fluid document formats in the age of ubiquitous computing. We are currently investigating different possibilities to support the versioning of fluid documents based on RSL's advanced linking and user rights management. Further research also has to be conducted on how to extend the document metaphor to deal with dynamic digital objects that are linked to physical components of surrounding environments, such as information retrieved from sensors. Finally, we are planning to extend our prototype for fluid document formats to deal with the adaptation of input as well as advanced output adaptation.

\vspace{0.3cm}

\bibliographystyle{ACM-Reference-Format}
\bibliography{main}

\end{document}